\documentclass{moriond}

\def\Journal#1#2#3#4<{{#1} {\bf #2}, #3 (#4)}

\def\be{\begin{equation}}
\def\ee{\end{equation}}
\def\bea{\begin{eqnarray}}
\def\eea{\end{eqnarray}}

\usepackage{amsmath,amssymb}

\newcommand{\Photo}{\includegraphics[height=35mm]{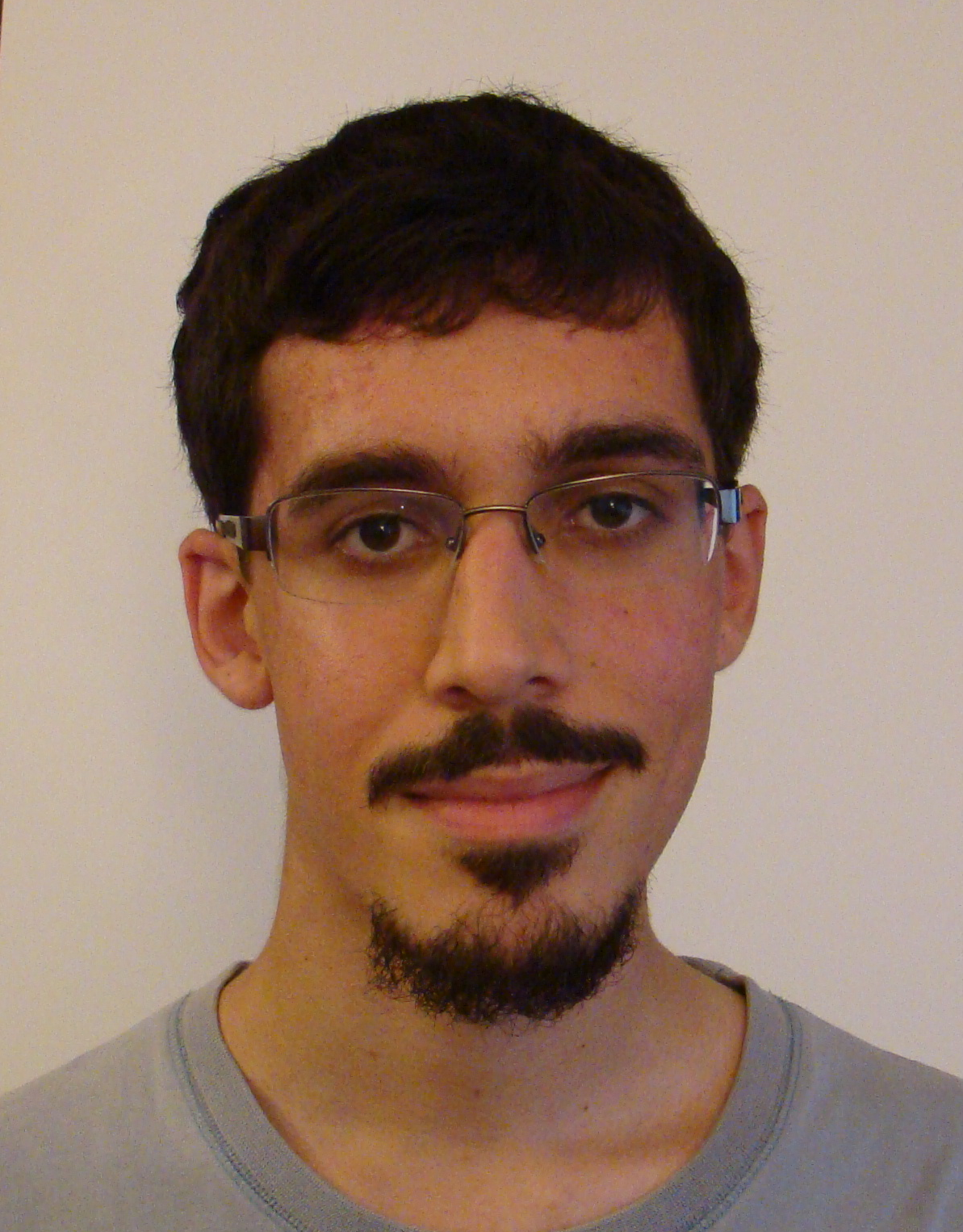}}

\begin{document}
\vspace*{4cm}
\title{European Spallation Source: a future for Coherent Neutrino Nucleus Scattering}

\author{Ivan Esteban}

\address{Center for Cosmology and AstroParticle Physics (CCAPP), Ohio State University, \\ Columbus, OH 43210}

\maketitle\abstracts{
The European Spallation Source (ESS), currently finishing its 
construction, will soon provide the most intense neutron beams for 
multi-disciplinary science. At the same time, it will also produce a 
high-intensity neutrino flux with an energy suitable for precision 
measurements of Coherent Elastic Neutrino-Nucleus Scattering. We 
describe some physics prospects, within and beyond the Standard Model, 
of employing innovative detector technologies to take the most out of 
this large flux. We show that, compared to current measurements, the 
ESS will provide a much more precise understanding of neutrino and 
nuclear properties.}

\section{Introduction}

In the Standard Model (SM), neutrinos only interact through the weak
interactions. This makes their properties sensitive to tentative
Beyond the Standard Model (BSM) effects with sub-weak strength. At the 
same time, they are a very clean probe to explore weak-interaction properties of 
different systems. Further pursuing this research program requires 
precision measurements of neutrino-matter interactions. 

In the SM, such interactions can take place between neutrinos and 
atomic nuclei through the weak neutral current, by exchanging a Z 
boson. Interestingly, if the momentum transfer $q$ remains smaller than 
the inverse nuclear size (which, for medium-sized nuclei, requires $|q| \lesssim 50 \, \mathrm{MeV}$), the process can take place coherently
with the whole nucleus. This dramatically enhances the interaction 
cross-section, that in the fully coherent regime is proportional to the 
square of the number of neutrons in the target nuclei.

Despite the large cross-sections that would open the way for large-statistics measurements of neutrino interactions, Coherent Elastic 
Neutrino-Nucleus Scattering (CE$\nu$NS) is very challenging to detect. The main reason is that the only observable final state is a recoiling nucleus with a kinetic energy in the few keV to sub-keV range. Indeed, CE$\nu$NS was only detected in 2017 by 
the COHERENT collaboration~\cite{Akimov:2017ade}, more than 40 years 
after its first theoretical description~\cite{freedman}.
This first detection, however, has opened up many research prospects, both within and 
beyond the SM~\cite{Kosmas:2017tsq,wma1,Cadeddu:2017etk,Coloma:2019mbs}. As present 
results are mostly limited by statistical uncertainties, high-statistics CE$\nu$NS measurements would undoubtedly broaden these prospects.

\section{CE$\texorpdfstring{\boldsymbol{\nu}}{v}$NS at the European Spallation Source}

\begin{figure}
\centering
\includegraphics[width=0.5\textwidth]{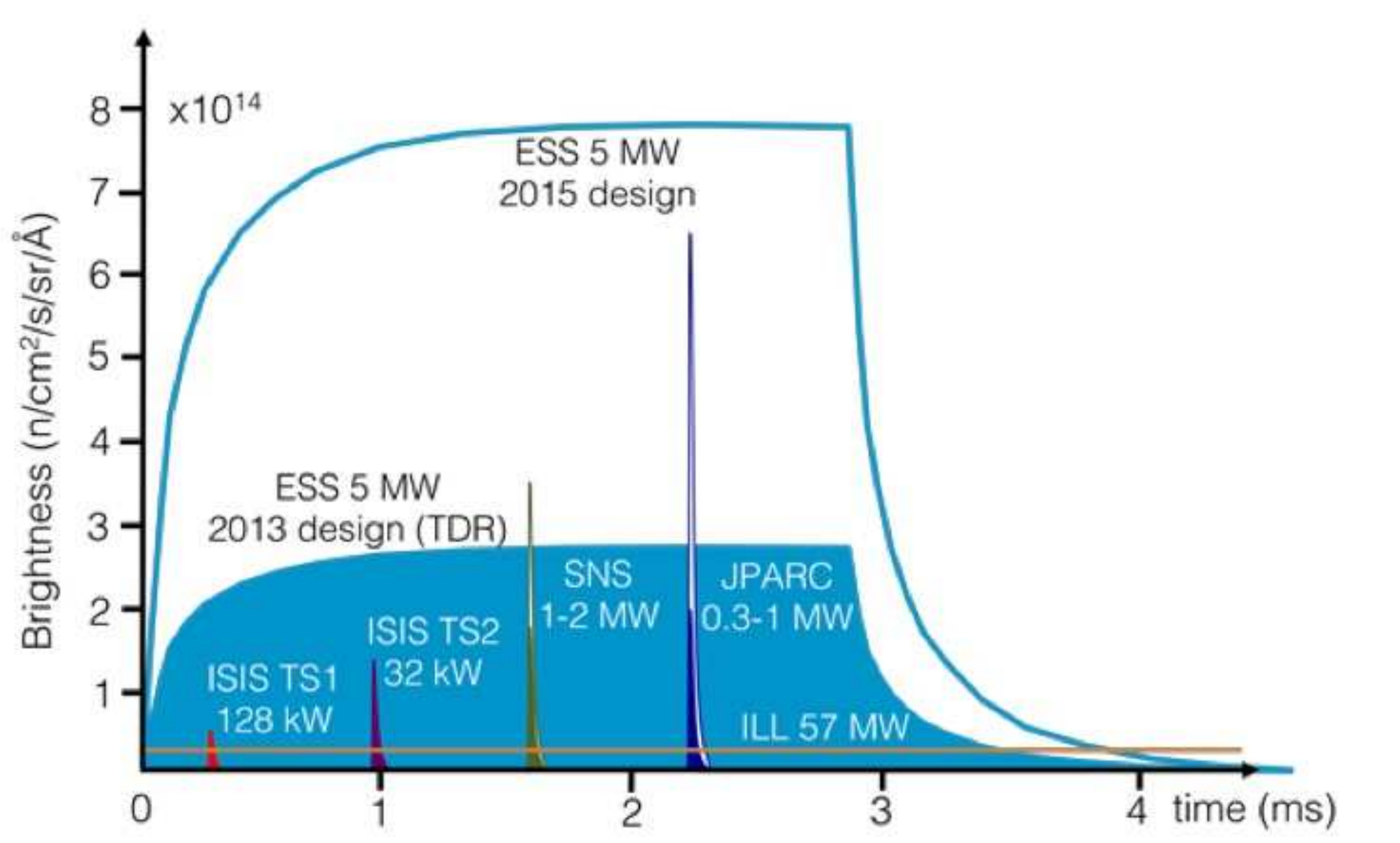}
\caption{Neutron production from existing and planned spallation sources, compared to the ESS. The COHERENT experiment is placed at the Oak Ridge Spallation Neutron Source, labelled as SNS in the figure.}
\label{fig:ess}
\end{figure}

There are two main strategies that would allow for high-statistics CE$\nu$NS measurements. On the one hand, using a more intense low-energy neutrino flux 
would increase the number of events while keeping the interaction 
coherent. A unique opportunity in this direction is brought by the 
European Spallation Source (ESS), whose user programme begins in 2023. 
It is a neutron spallation source, that produces a well-understood 
neutrino flux through pion decay at rest. This same operating principle 
generates the neutrinos used at COHERENT, but at the ESS the 
neutrino flux should be larger by about one order of magnitude~\cite{Baxter:2019mcx} (see Fig.~\ref{fig:ess}).

\begin{figure}
\centering
\includegraphics[width=0.5\textwidth]{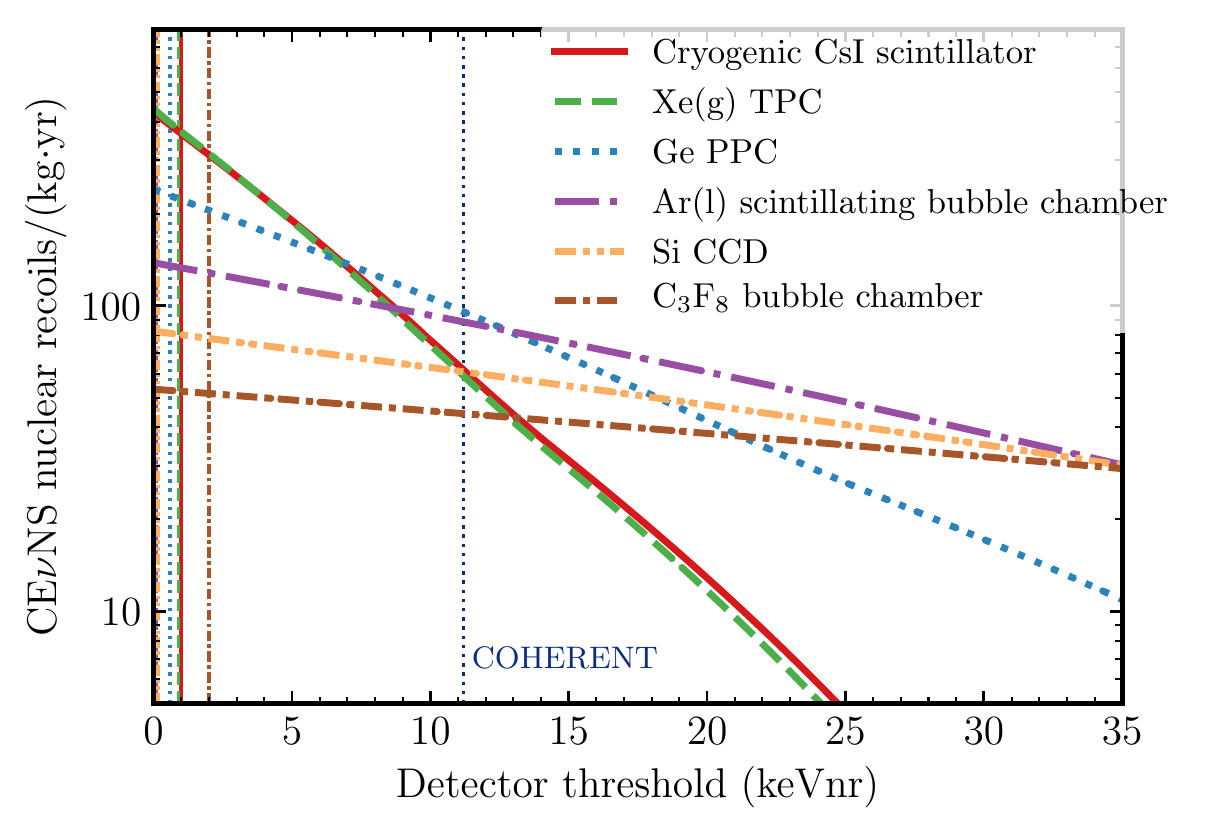}
\caption{Expected integrated CE$\nu$NS rate above nuclear recoil threshold, 20 m away from the ESS target, for different detectors. The vertical lines show the thresholds considered in Ref.~\protect\cite{Baxter:2019mcx}, and for illustration we also show in blue the threshold of the COHERENT CsI detector. See Ref.~\protect\cite{Baxter:2019mcx} for more details.}
\label{fig:threshold}
\end{figure}

On the other hand, the CE$\nu$NS cross-section strongly increases for 
small nuclear recoil energies. Thus, employing state of the art, low-threshold detectors would also enhance the statistical 
power of CE$\nu$NS measurements. This is illustrated in Fig.~\ref{fig:threshold}, that shows the expected number of events at the ESS as a function of the detector energy threshold for different materials. As the ESS is still under construction, suitable space can be allocated for such detectors~\cite{Baxter:2019mcx}.

\section{Physics prospects}
As an example of the physics reach of the strategies discussed above, we consider two scenarios characteristic of BSM and SM physics that can be probed with CE$\nu$NS: non-standard 
neutrino interactions (NSI) and neutron radii of nuclei. Further 
scenarios are explored in Refs.~\cite{Baxter:2019mcx,EstevesChaves:2021jct}.

From a model-independent approach, a useful parametrization of BSM effects is through the addition of higher-dimensional operators to the SM Lagrangian. At dimension 6, the allowed set of operators includes four-fermion operators affecting neutrino production, propagation and detection. They include the so-called neutral current vector NSI
\begin{equation}
2\sqrt{2} G_F \varepsilon_{\alpha \beta}^{f, V} (\bar{\nu}_\alpha \gamma_\mu P_L \nu_{\beta}) (\bar{f} \gamma^\mu f) \, ,
\end{equation}
where $G_F$ is the Fermi constant, $\alpha$ and $\beta$ are neutrino flavors, $f$ is a SM fermion, 
and $P_L$ is a left-handed projection operator. These operators are very challenging to 
constrain, due to the uncertainties in computing high-energy neutrino-nucleus interactions and the experimental difficulties in measuring 
neutral current cross sections precisely. Furthermore, they can 
significantly impact the interpretation of neutrino oscillation data~\cite{nsi10,Bakhti:2014pva,Esteban:2020itz}. 

\begin{figure}
\centering
\includegraphics[width=\textwidth]{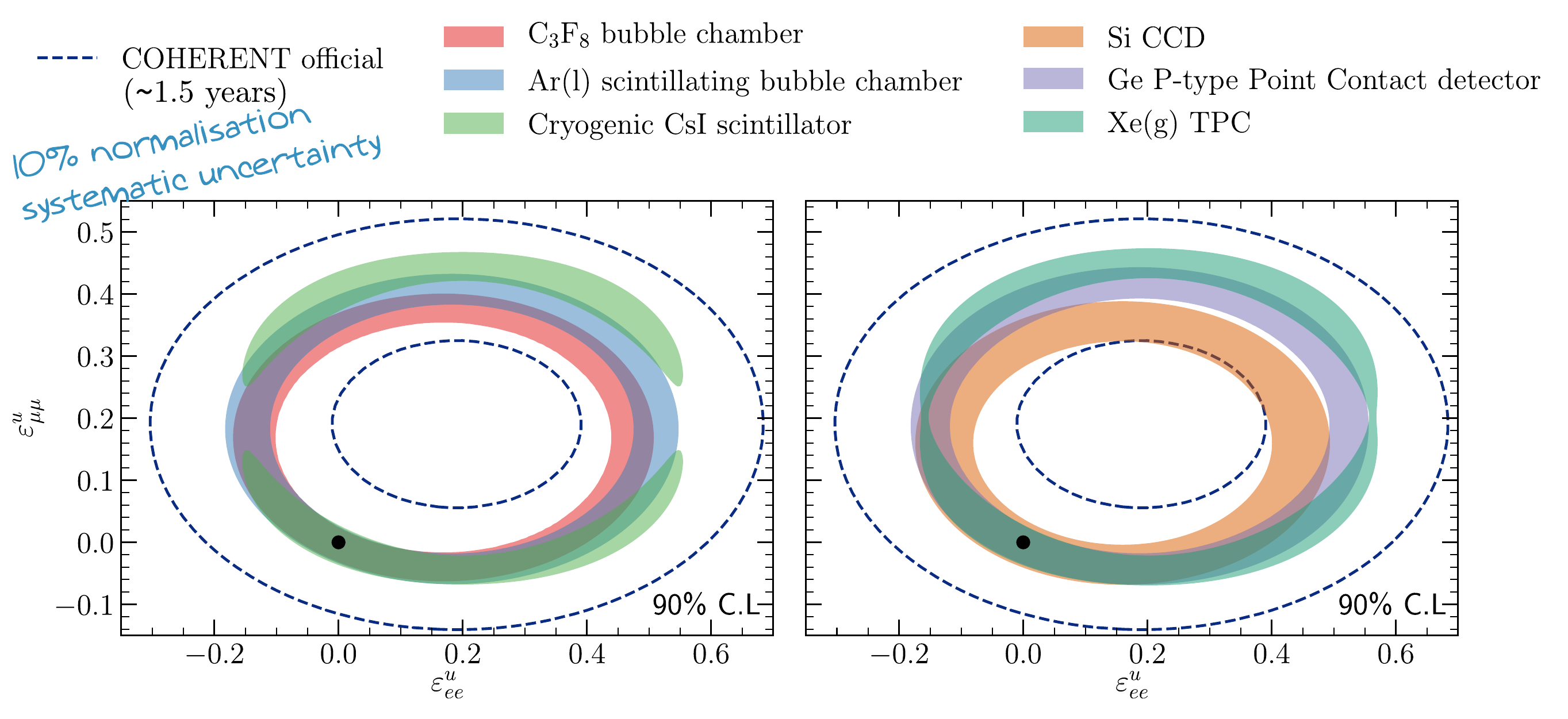}
\caption{Expected allowed regions for vector NSI with up quarks in the ($\varepsilon_{ee}^{u, V}$, $\varepsilon_{\mu\mu}^{u, V}$) at the 90\% confidence level, after 3 years of data taking at the ESS for different detector materials. For simplicity, the rest of the NSI parameters not shown in the figure have been set to zero. See Ref.~\protect\cite{Baxter:2019mcx} for more details.}
\label{fig:NSI2D}
\end{figure}

These same operators, however, also affect CE$\nu$NS. Thus, this process can provide competitive constraints on NSI, completely independently from neutrino oscillation bounds. Figure~\ref{fig:NSI2D} shows the 90\% confidence level bounds that could be obtained after 3
years of data taking at the ESS with different detector materials. These would significantly tighten present constraints.

An example of physics within the SM that can be probed with CE$\nu$NS
is the neutron radius of nuclei, a quantity for which almost no model-independent measurements exist. CE$\nu$NS is a process that takes 
place essentially with the neutrons in the target nucleus, and it is
therefore sensitive to their distribution inside it. This is precious 
information to complement proton densities accessible with elastic 
electron scattering, and its understanding impacts the limits of 
existence~\cite{Erler2012} and size~\cite{Tanihata:2013jwa} of atomic nuclei, as well as being an important test 
of first-principles nuclear calculations~\cite{Tsang:2012se}. Beyond the structure of 
nuclei, neutron distributions can be linked to the properties of 
neutron-rich matter, which determines the size and structure of neutron 
stars~\cite{Lattimer:2004pg,Lattimer:2006xb}.

\begin{figure}
\includegraphics[width=\textwidth]{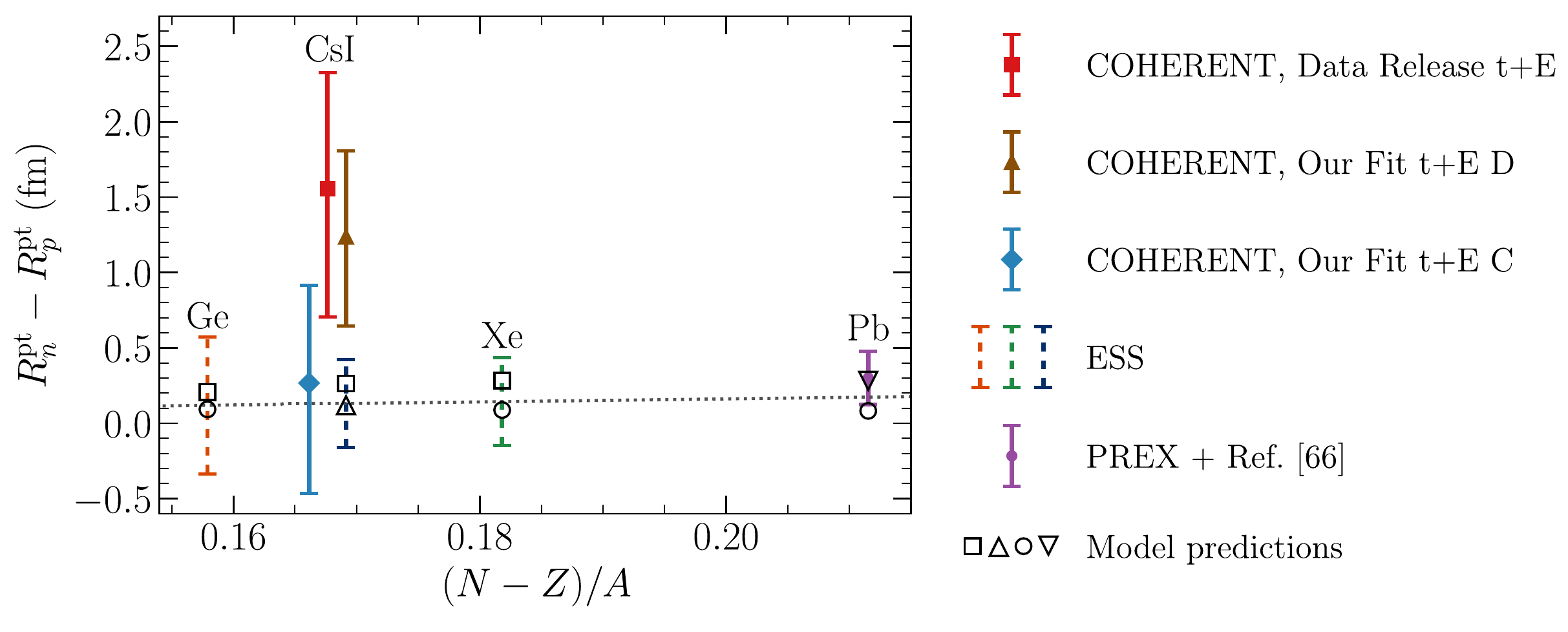}
\caption{Present determination and future sensitivity of the neutron skin thickness from CE$\nu$NS experiments for different nuclei.  See Ref.~\protect\cite{Coloma:2020nhf} for more details.}
\label{fig:Rn}
\end{figure}

Figure~\ref{fig:Rn} shows the sensitivity that the ESS can achieve on 
the neutron radius of various nuclei, compared to the present COHERENT sensitivity and 
to different theoretical nuclear physics model predictions. As the results show, 
high-statistics measurements at the ESS would have an experimental 
sensitivity of the order of the spread among the theoretical models. 

\section*{References}

\bibliography{Bibliography}
\end{document}